\documentclass[a4paper]{jpconf}

\usepackage{graphicx}
\usepackage{subscript}

\begin{document}

\title{Development of microwave-multiplexed superconductive detectors for the HOLMES experiment}
\author{A.~Giachero$^1$~, 
        D.~Becker$^3$~, 
        D.A.~Bennett$^3$~, 
        M.~Faverzani$^{1,2}$~, 
        E.~Ferri$^{1,2}$~, 
        J.W.~Fowler$^3$~, 
        J.D.~Gard$^3$~, 
        J.P.~Hays-Wehle$^{1,3}$~, 
        G.C.~Hilton$^3$~, 
        M.~Maino$^1$~, 
        J.A.B~Mates$^3$~, 
        A.~Puiu$^{1,2}$~, 
        A.~Nucciotti$^{1,2}$~,
        C.D.~Reintsema$^3$~, 
        D.R.~Schmidt$^3$~, 
        D.S.~Swetz$^3$~, 
        J.N.~Ullom$^3$~, 
        L.R~Vale$^3$}

\address{$^1$INFN of Milano-Bicocca, Department of Physics, Milan, Italy}
\address{$^2$University of Milano-Bicocca, Department of Physics, Milan, Italy}
\address{$^3$National Institute of Standards and Technology, Boulder, CO, USA}

\ead{andrea.giachero@mib.infn.it}

\begin{abstract}
  In recent years, the progress on low temperature detector technologies has allowed design of large scale experiments aiming at pushing down the sensitivity on the neutrino mass below 1\,eV. Even with outstanding performances in both energy ($\sim$eV on keV) and time resolution ($\sim 1\,\mu$s) on the single channel, a large number of detectors working in parallel is required to reach a sub-eV sensitivity. HOLMES is a new experiment to directly measure the neutrino mass with a sensitivity as low as 2\,eV. HOLMES will perform a calorimetric measurement of the energy released in the electron capture (EC) decay of \textsuperscript{163}Ho. In its final configuration, HOLMES will deploy 1000 detectors of low temperature microcalorimeters with implanted \textsuperscript{163}Ho nuclei. The baseline sensors for HOLMES are Mo/Cu TESs (Transition Edge Sensors) on SiN\textsubscript{x} membrane with gold absorbers. The readout is based on the use of rf-SQUIDs as input devices with flux ramp modulation for linearization purposes; the rf-SQUID is then coupled to a superconducting lambda/4-wave resonator in the GHz range, and the modulated signal is finally read out using the homodyne technique. The TES detectors have been designed with the aim of achieving an energy resolution of a few eV at the spectrum endpoint and a time resolution of a few micro-seconds, in order to minimize pile-up artifacts.

\end{abstract}

\section{Introduction}

The measurement of the end point of nuclear beta or electron capture (EC) decays spectra is the only theory-unrelated method usable to sort out the problem of the neutrino mass. Among the family of nuclear beta or electron capture decays, an interesting isotope suitable for the neutrino mass experiment is the \textsuperscript{163}Ho, which decays via EC with $Q_{EC}<3$~keV, as proposed by De Rujula and Lusignoli in 1982~\cite{DeRujula}. The HOLMES experiment~\cite{HOLMES} will deploy a large array of low temperature microcalorimeters with implanted \textsuperscript{163}Ho nuclei. The detectors used for the HOLMES experiment will be Mo/Cu TESs (Transition Edge Sensors) on SiN\textsubscript{x} membrane with gold absorbers. The TES microcalorimeters will be fabricated in a two step process. The first step is the fabrication of the detectors themselves, to be carried out at the National Institute for Standard and Technology (NIST, Boulder, Co, USA). The second step is the \textsuperscript{163}Ho nuclei implantation into the calorimeter absorber, to be carried out at the INFN laboratories in Italy. The TES design will be optimized to have $\Delta E \simeq 1$\,eV at the \textsuperscript{163}Ho EC endpoint and a resolving time of $\tau_r\simeq 1$\,$\mu$s to reject pile up events at the expected event rate (300\,Hz/pixel). Microwave frequency domain read out is the best available technique to read out large array of such detectors. In this way, a thousand channels can be multiplexed onto the same coaxial line, and readout thereof is limited only by the the bandwidth of the available commercial fast digitizers. This microwave multiplexing system will be used to read out the HOLMES detectors. 

\begin{figure}[!t]
  \begin{center}
    \includegraphics[clip=true,width=\textwidth]{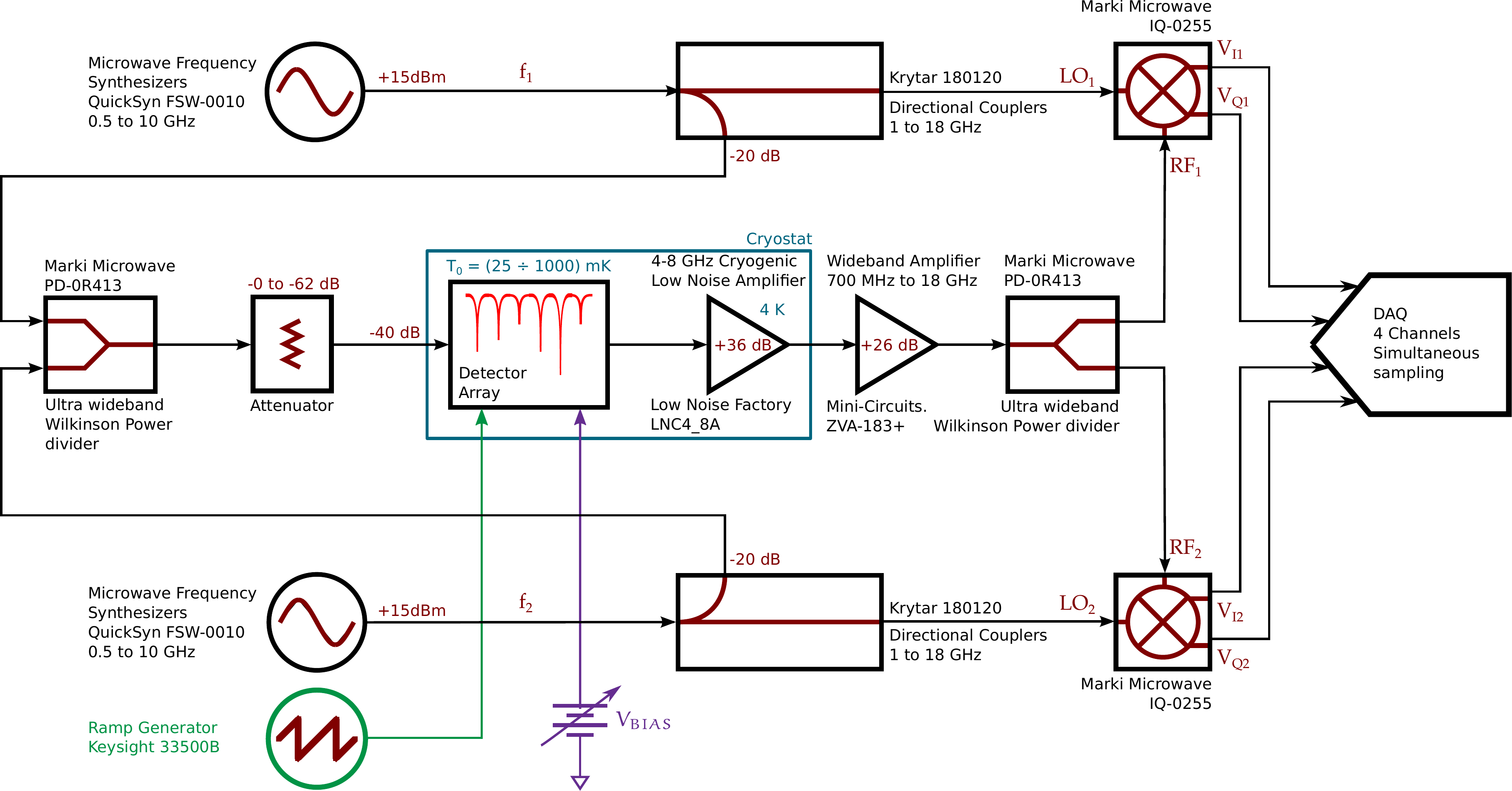}
  \end{center}
  \caption{\label{fig:2ch-setup} Block diagram of the developed two-channel readout and multipling system.}
\end{figure}

\section{Setup and Multiplexing systems}
Multiplexed readout is an essential requirement for kilo-pixel arrays. To date, many techniques have been developed to multiplex TES arrays with SQUID readout: Time Division (TDM)~\cite{TDM}, Frequency Division (FDM)~\cite{FDM} and Code Division (CDM)~\cite{CDM}. However, the scalability of these readout approaches is limited by the finite measurement bandwidth ($\simeq$10\,MHz) achievable in a flux-locked loop. Other sensors, like Microwave Kinetic Inductance Detectors (MKIDs), are naturally frequency-multiplexed and provide a higher multiplexing factor, but with performances in X-ray spectroscopy not yet competitive with other superconducting detectors. 

Microwave SQUID multiplexing ($\mu$Mux)~\cite{RFSQUID} is a readout technique that potentially combines the proven sensitivity of TESs and the scalable multiplexing power found in Microwave Kinetic Inductance Detectors (MKIDs). This method uses radio-frequency (rf) SQUIDs as input devices with flux ramp modulation, coupled to high quality-factor (Q) microwave superconducting lambda/4-wave microresonators in the GHz range. Absorbed events in the microcalorimeter cause a temperature increase. Since the TES edge is sharp, a tiny change in temperature results as large resistance ($R_{TES}$) increase, that causes a decrease in current that flows inside the sensor ($I_{TES}$). The current variation is transduced in a change in the input flux to the SQUID that causes a change of resonance frequency and phase of the superconducting microresonator. By acquiring the in-phase (I) and quadrature (Q) signals, it is possible to recover the changes in the resonance frequency and phase of each microresonator. The ramp induces a controlled flux variation in the rf-SQUID, which is crucial for linearizing the response~\cite{RAMP}.

By tuning the lambda/4-wave resonators at different frequencies it is straightforward to multiplex many RF carriers. The comb of frequency carriers is generated by digital synthesis in the MHz range and up-converted to the GHz range by an IQ-mixer. The GHz range comb is sent to the cold $\mu$MUX chips coupled to the TES array through one semi-rigid cryogenic coax cable, amplified by a cryogenic low noise High Electron Mobility Transistor (HEMT) amplifier, and sent back to room temperature through another coax cable. The output signal is down-converted by a second IQ-mixer, sampled with a fast A/D converter, and digital mixing techniques are used to recover the signals of each TES in the array (channelization). The number of TES signals which can be multiplexed is set by the bandwidths of both the A/D converter and the TES signal. The final version of the HOLMES readout system will be based on an open source reconfigurable open architecture computing hardware named ROACH2 (Reconfigurable Open Architecture Computing Hardware) equipped with a FPGA Xilinx Virtex6~\cite{ARCONSReadout}. 

The development of the HOLMES multiplexing and read-out system started in 2015 and will be ready in the 2016. In the meanwhile a two-channel system, based on commercial components, has been developed. The goal of this development is to have an independent system ready-to-use for the detector characterizations and tests. The block diagram is shown  in figure \ref{fig:2ch-setup}. Two different probe signals $LO_1$ and $LO_2$, one for each microresonator, are generated by two microwave synthesizers and then merged together by a power coupler. The resulting signal is used to excite the resonators array. The forward transmitted waveform is amplified with a cryogenic HEMT amplifier, mounted on the 4\,K stage, and with a room-temperature amplifier. The amplified signal is then split by a power divider in two copies, $RF_1$ and $RF_2$. Homodyne mixing between each copy ($RF_1$ and $RF_2$) and the original excitation signals ($LO_1$ and $LO_2$) is accomplished employing two IQ-mixers. The resulting $V_{Q_i}$ and $V_{I_i}$ signals are acquired by a commercial 14bit-simultaneous-sampling data acquisition board (NI PXI 6133) with a sample rate of $f_s=2.5$\,MHz. 


\begin{figure}[!t]
\includegraphics[clip=true,width=0.5\textwidth]{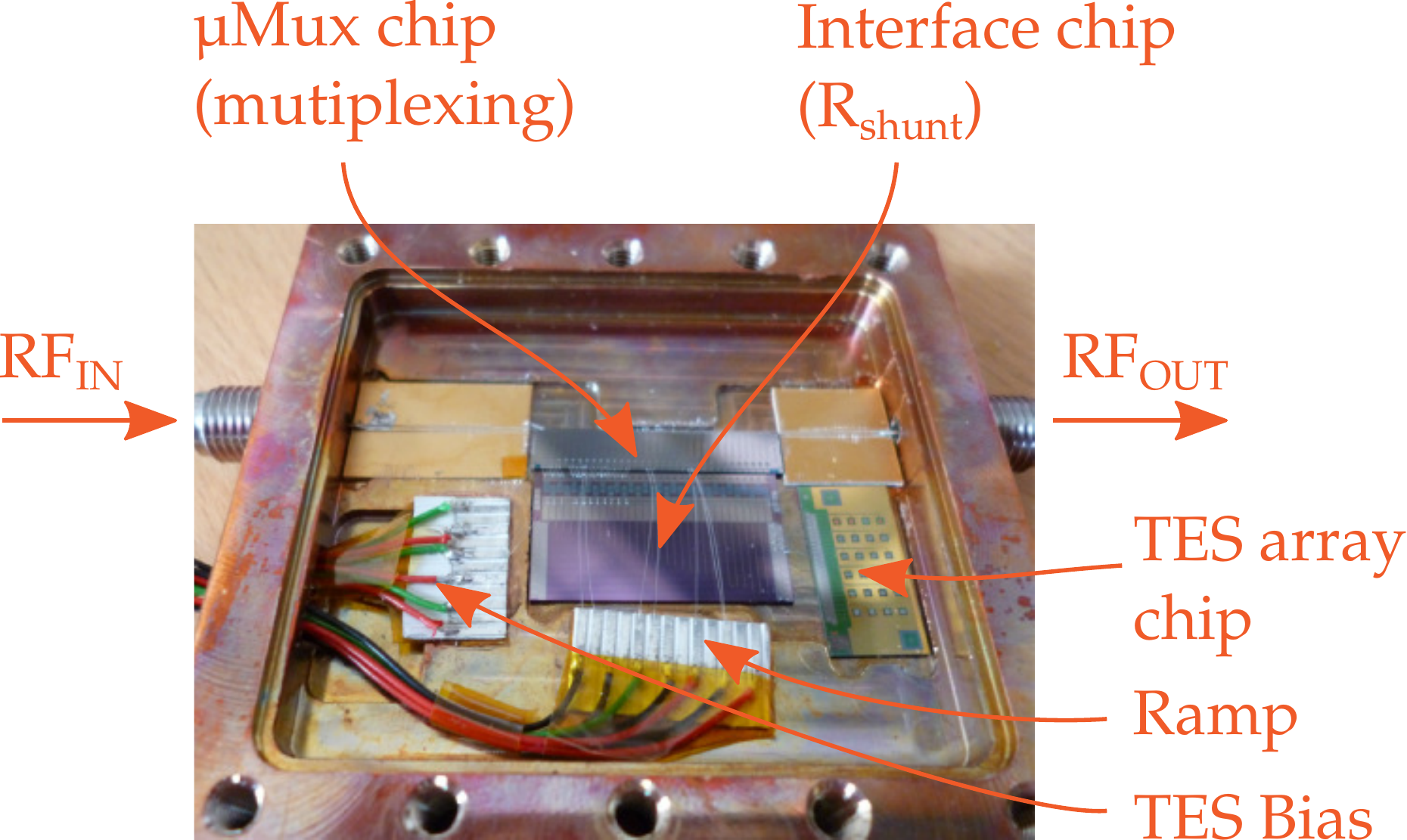}\hspace{2pc}%
\begin{minipage}[b]{14pc}\caption{\label{fig:setup}Photograph of the microwave rf-SQUID set-up and X-ray TESs with Bismuth absorbers. TESs are wirebonded to rf-SQUID input coils on the multiplexer ($\mu$Mux) chip through a central interface chip.}
\end{minipage}
\end{figure}

The ramp is generated by using a remote programmable waveform generator (Keysight 33500B), with a frequency in the range of 40 to 100\,kHz. The amplitude of the ramp is tuned such that it provides $\sim 3\Phi_0$ of flux per ramp period and modulates $V_{Q_i}$ and $V_{I_i}$ at a carrier frequency of $f_c \simeq (120\div 300)$\,kHz. In order to avoid noise and pick-up due to ground loops the ramp line and the ramp generator are isolated by a unitary isolation amplifier powered by batteries. The TES bias is provided by a custom box composed by a programmable power supply (Agilent 3631A) and a voltage divider in order to have a $I_{BIAS}$ within the range $(0 - 1000)\,\mu$A. Reading out the voltage across a 1,k$\Omega$ resistor placed in series at the divider output it is possible to measure the current $I_{BIAS}$ provided to the TES. The power supply can be replaced by a battery with voltage tuned by a trimmer that forms a voltage divider. The programmable solution is used during the I-V curves measurements while the battery solution is used during the normal data taking in order to avoid noise and pick-up.

The microcalorimeter TES detector array consists of $4 \times 6$ pixels fabricated at NIST (Boulder, CO, USA). The superconducting thin film is a Molybdenum-Copper bilayer film, where the proximity effect~\cite{Proximity} is used to achieve a critical temperature of $T_c\simeq 100$\,mK. The TESs are suspended on a silicon-nitride (Si\textsubscript{3}N\textsubscript{4}) membrane $1\,\mu$m thick for thermal insulation. The absorber is made of bismuth with horizontal dimensions of $350 \times 350\,\mu$m and with $2.5\,\mu$m thickness. This TES array is optimized for X-ray spectroscopy but not specifically designed for HOLMES. An interface chip is used to provide a bias shunt resistance of $R_{shunt} = 0.33$\,m$\Omega$ in parallel with each TES and a wirebond-selectable Nyquist inductor, $L_{N}$, in series to tune the pulse rise-times. The multiplexer chip ($\mu$Mux) consists of 33 quarter-wave coplanar waveguide (CPW) microwave resonators made from a 200\,nm thick Nb film deposited on high-resistivity silicon ($\rho>10$\,k$\Omega\cdot$cm). Adjacent resonances are separated by (2-6)\,MHz and are centered around 5.72 GHz. In our tests we measured four pixels with resonance frequencies at 5.6338\,GHz, 5.6341\,GHz, 5.6455\,GHz and  5.6711\,GHz respectively. All the chips are placed inside a copper holder connectorized with two SMAs for the feedline and a Micro-D for the bias and ramp lines. An X-ray source (\textsuperscript{55}Fe) is mounted externally to the holder and it irradiates the sensitive area of the detectors by an opening on top side covered by a aluminum foil, 12~$\mu$m thick, in order to avoid noise due to black-body radiation. The holder is cooled down in the range $(40-90)$\,mK by a pulse tube assisted dilution cryostat (Oxford Triton 200), the same that will host the HOLMES experiment in its final configuration. 

\begin{figure}[!t]
\includegraphics[clip=true,width=0.5\textwidth]{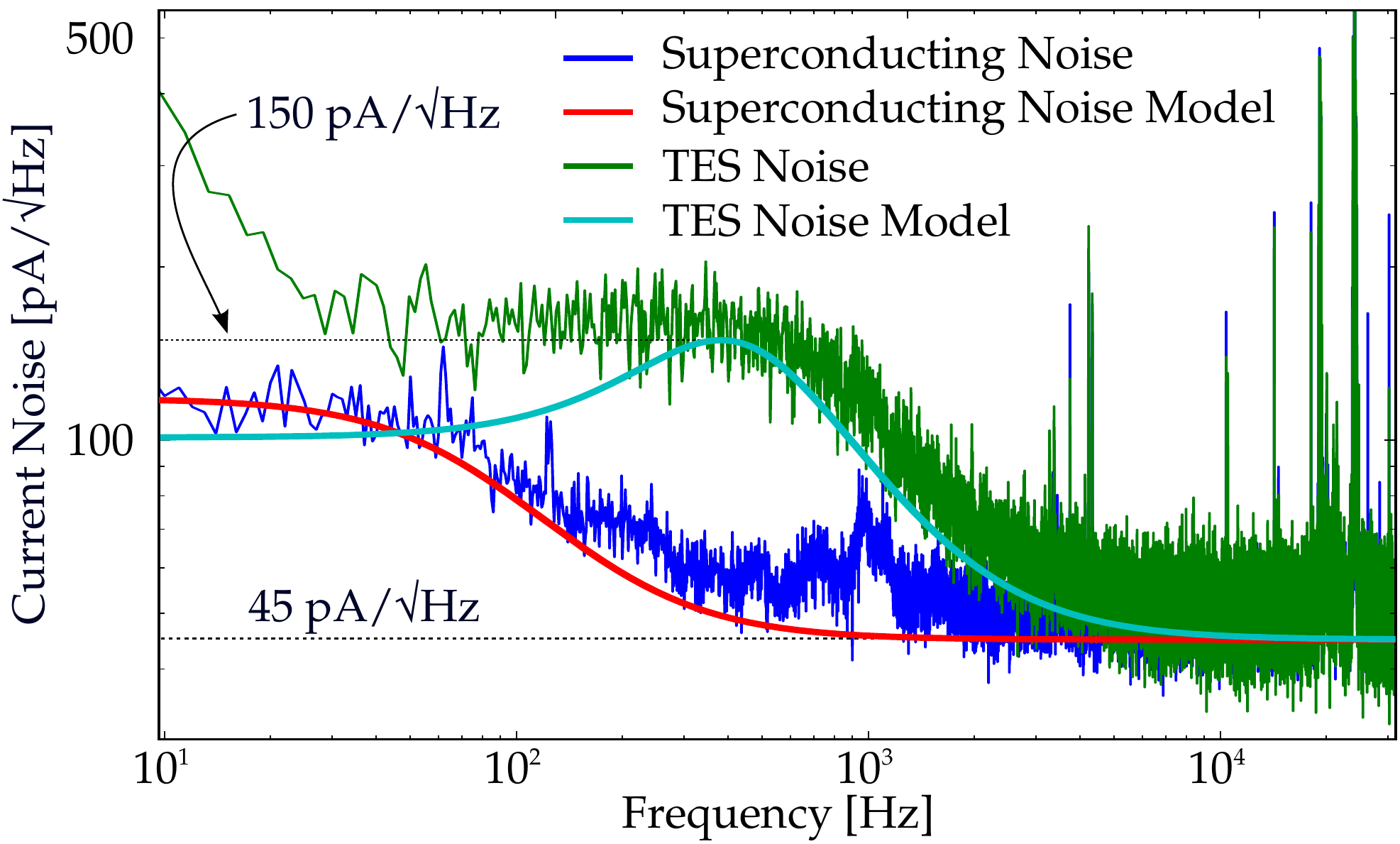}\hspace{2pc}%
\begin{minipage}[b]{14pc}\caption{\label{fig:noise} Experimental noise spectra for biased and unbiased TES together with the theoretical model.}
\end{minipage}
\end{figure}


\section{Measurements results}
The goal of these first measurements is the development of a microwave read out system with optimized performances for the HOLMES detectors characterization. To achieve this goal we are using the aforementioned TES array whose behavior and response are well-known. Results obtained with developed system can be compared with theoretical models and with previous characterization performed at NIST with different read-out techniques. 

\begin{figure}[!t]
  \begin{center}
    \includegraphics[clip=true,width=0.9\textwidth]{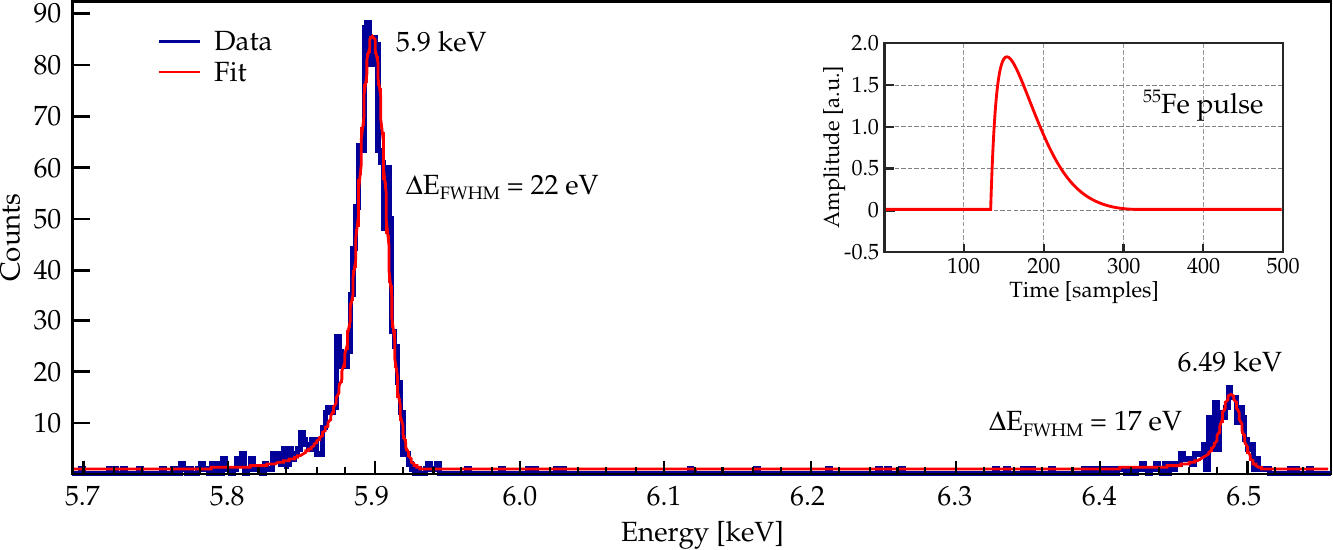}
  \end{center}
  \caption{\label{fig:spectra} First energy spectrum from an \textsuperscript{55}Fe X-ray source obtained by the developed microwave readout system. Inlet: example of triggered pulse.}
\end{figure}

The DAQ system and the ramp generator run with the same 10\,MHz clock reference. The data acquisition is synchronized with the ramp by the ramp generator synch probe. The I and Q data are demodulated from the flux-ramp-modulated signal by measuring the phase shift~\cite{RAMP}. The demodulated signal are triggered by a 2\textsuperscript{nd}-order derivative algorithm and filtered offline by optimum filter based techniques. The TESs were biased at 20\% of their normal state resistance ( $R_N$). In order to estimate the achievable resolution we evaluated the system total noise (figure \ref{fig:noise}). We also compared the obtained spectra with that which is expected from the theoretical model computed by using the Irwin-Hilton theory~\cite{TES}, for a working temperature of 65\,mK. The obtained noise agrees with the model except for high frequency peaks due to pick up and a low frequency $1/f$ component due to slow thermal drifts of the mixing chamber stage. Fortunately, both of these effects are outside of the important signal band. Nevertheless, we are working on mitigating them with better shielding and an improved temperature regulation system,

The first obtained spectra is show in figure \ref{fig:spectra}. The two peaks (5.9\,keV and 6.42\,keV) deviate from Gaussian presenting a low energy tail on the spectral lines. This is a known behavior for Bismuth absorber probably due to thermal phonon losses. To consider this deviation the peaks are fitted by using a Gaussian core portion with a power-law low-end tail. Considering the signal-to-noise ratio around 1000, the expected resolution at 6\,keV results around $(2 -5)$\,eV. A preliminary analysis showed a resolution of 22\,eV~@~5.9\,keV and 17\,eV~@~6.49\,keV, an order of magnitude worse. A possible explanation for this is the following. In order to match the current ADC bandwidth we slowed down the detector by tuning the wirebond-selectable Nyquist inductor $L_N$ on the TES bias circuit. With slower detectors, the thermal regulation of the cryostat becomes very important, as thermal drifts can occur on the timescale of the pulse, giving a shape to the baseline that cannot be easily compensated. We intend not only to improve the temperature stability of the mixing chamber stage, as mentioned above, but also to speed up the detector pulse such that it will occur over a time period short enough that the baseline can be treated as constant. In order to verify this hypothesis we are developing a new read-out system based on a faster ADC board ($f_s=250$\,MS/s).

\section{Future developments}
In this work we have demonstrated that microwave rf-SQUID multiplexing readout is an excellent candidate for X-ray spectroscopy. In the current setup, in order to match the current ADC bandwidth we slowed down the detector by tuning the Nyquist inductor. The goal was to test and demostrate the technique. Currently we are developing a new read-out system based on a faster ADC board ($f_s=250$\,MS/s). This system will fit with the needed requirements and it will be used to evaluate all the relevant parameters (energy and time resolutions, thermal capacity, ...) of the incoming first HOLMES detector prototypes.

\section*{Acknowledgements}
This work was supported by the European Research Council (FP7/2007-2013) under Grant Agreement HOLMES no. 340321. We also acknowledge the support from INFN through the MARE project and from the NIST Innovations in Measurement Science program for the TES detector development.

\section*{References}
\bibliography{giachero-TAUP2015}

\end{document}